\documentclass[pra,twocolumn,groupedaddress,showpacs,floatfix]{revtex4}

\usepackage{graphics}
\usepackage{bm}
\usepackage{amsmath}
\usepackage{amsfonts}
\usepackage{amssymb}
\usepackage{latexsym}

\begin{document}

\title{Entangled state preparation via dissipation-assisted adiabatic passages} 
\author{Carsten Marr$^1$}
\author{Almut Beige$^{1,2}$}
\author{Gerhard Rempe$^1$}
\affiliation{$^1$Max-Planck-Institut f\"ur Quantenoptik, Hans-Kopfermann-Str. 1,  85748 Garching, Germany}
\affiliation{$^2$Blackett Laboratory, Imperial College London, Prince Consort Road, London SW7 2BZ, UK}

\date{\today}

\begin{abstract}
The main obstacle for coherent control of open quantum systems is decoherence due to different dissipation channels and the inability to precisely control experimental parameters. To overcome these problems we propose to use {\em dissipation-assisted adiabatic passages}. These are relatively fast processes where the presence of spontaneous decay rates corrects for errors due to non-adiabaticity while the system remains in a decoherence-free state and behaves as predicted for an adiabatic passage. As a concrete example we present a scheme to entangle atoms by moving them in and out of an optical cavity. 
\end{abstract}

\maketitle

\section{Introduction}

In recent years several schemes to entangle atoms \cite{plenio,beige4,foeldi,sorensen} and to implement gates for quantum information processing \cite{beige2,beige3,zheng,pachos,jane,you} using optical cavities have been proposed. Cold atoms trapped in an optical cavity provide a promising technology for quantum  computing as well as an ideal model for theoretical studies. The main problems that must be overcome, decoherence due to different dissipation channels and the inability to precisely control experimental parameters, are common with other potential implementations. Dissipation results from the fact that the atom-cavity coupling constant $g$ is of about the same size as the spontaneous photon and atom decay rates $\kappa$ and $\Gamma$. Optical cavities operate in a parameter regime with
\begin{eqnarray} \label{aim}
g \sim \kappa \sim \Gamma ~.
\end{eqnarray}
Since it takes at least the time $1/g$ to create a significant amount of  entanglement between the atoms, it seemed impossible to avoid spontaneous emission and the loss of information stored in the system. 

Some of the proposed schemes work only with a success rate below $50\,\%$ \cite{plenio,sorensen}. Others try to solve the dissipation problem by avoiding the population of excited states with the help of adiabatic population transfers between ground states and strongly detuned laser fields \cite{foeldi,zheng,jane,you} or use the existence of decoherence-free states and an environment-induced quantum Zeno effect \cite{beige4,beige2,beige3,pachos}. While many of these schemes are able to suppress one type of dissipation very well, their operation time is much longer than the inverse atom-cavity coupling constant. Thus, while for example the probability for leakage of photons is made very small, failure of the proposed scheme becomes inevitable due to spontaneous emission from the atoms. As regards dissipation, the quantum computing scheme proposed by Pellizzari {\em et al.}  in 1995 \cite{pellizzari} is still one of the most efficient. Demanding a certain minimum fidelity and success rate, it requires a relatively small ratio between $g^2$ and $\kappa \Gamma$ and is nearly comparable to the schemes proposed in \cite{pachos,you}.

\begin{figure}
\begin{minipage}{\columnwidth}
\begin{center}
\resizebox{\columnwidth}{!}{\rotatebox{0}{\includegraphics{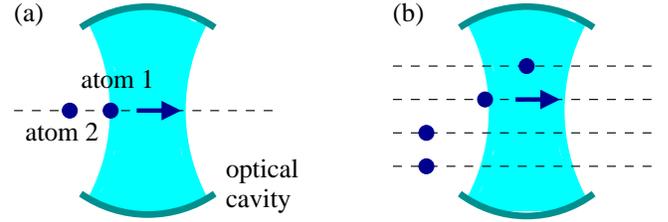}}}  
\end{center}
\vspace*{-0.2cm}
\caption{Entangled state preparation. Two atoms can be prepared in a maximally entangled state by moving them slowly into an optical cavity (a). More general, entangled states of up to $N$ atoms are prepared by moving them together or separately in and out of the resonator (b).} \label{setup}
\end{minipage}
\end{figure}

In this paper an alternative way to entangle atoms with the help of an optical cavity is proposed (see Figure \ref{setup}). Relatively short operation times are achieved by using {\em dissipation-assisted adiabatic passages} \cite{simple}. In the following, the probability for no photon emission and success rate of the proposed schemes is for a wide range of parameters above $90 \, \%$ while the fidelity of the prepared state can be of the order of $99 \,\%$. In addition, the experimental requirements for coherent state preparation are significantly reduced. The scheme is robust against fluctuations of most system parameters; it does not require cooling of the atoms into the Lamb-Dicke regime  \cite{kimble} neither demands precise control of the atom-cavity coupling constants. Moreover, the scheme does not involve individual laser addressing of the atoms inside the cavity.

To avoid errors due to leakage of photons through the cavity mirrors, the system should remain during the whole state preparation in a decoherence-free (DF) state \cite{palma,zanardi,lidar} with respect to this dissipation channel. Decoherence-free states are states whose population does not lead to a photon emission. A system of $N$ two-level atoms inside an optical cavity possesses a DF subspace of dimension ${N \choose (N+1)/2}$ or ${N \choose N/2}$ for odd and even numbers of atoms, respectively \cite{beige4}.

The central idea of the proposed scheme is to manipulate the system by slowly changing the atom-cavity coupling constants which define the DF states of the system. Initially prepared in a DF state, the system remains DF and follows the parameter change adiabatically \cite{facchi}. This is a consequence of the adiabatic theorem \cite{adia} and relies on the fact that the DF states of the system are at the same time the eigenstates of the atom-cavity interaction Hamiltonian. Here we consider evolutions where the initial state of the atoms is a product state, while the final state is highly entangled, providing a simple and efficient atom entangling scheme. The easiest way to vary the atom-cavity coupling constants is by moving the atoms through the cavity as shown in Figure \ref{setup}. That this is feasible with present technology has already been demonstrated \cite{rempe,guthoerlein}. Alternatively, the atoms might be trapped in an optical lattice above an atom chip and move through a micro cavity installed on the surface of the chip \cite{horak}. 

In the following it is shown that cavity decay can stabilise the desired time evolution of the system. It increases the fidelity of the prepared state under the condition of no photon emission if the scheme is operated relatively fast, i.e. outside the adiabatic regime. This is achieved since population that accumulates due to non-adiabaticity in unwanted states is damped away during the no-photon time evolution. Under the condition of no emission, the system behaves as predicted by the adiabatic theorem and the resulting time evolution can be called a dissipation-assisted adiabatic passage. The overall operation time can be reduced by as much as two orders of magnitude. To a good approximation, the presence of the cavity decay rate has the same effect as an error detection measurement. 

A detailed description of a scheme to prepare two atoms in a maximally entangled state and its underlying structure are given in Section \ref{sectio}. Section \ref{section} discusses the state preparation scheme from an experimental point of view and introduces variations of the proposed experiment to increase its feasibility. Finally, we summarise our results in Section \ref{conc} and point out potential applications of dissipation-assisted adiabatic passages for quantum state preparation of more than two atoms and in quantum information processing.

\section{Dissipation-assisted state preparation} \label{sectio}

To illustrate the basic idea underlying the state preparation schemes discussed in this paper, we give in this section a detailed analysis of a simple process to transfer two atoms with ground state $|1 \rangle$ and excited state $|2 \rangle$ into the maximally entangled state 
\begin{eqnarray} \label{a}
|a \rangle & \equiv & {1 \over \sqrt{2}} \, \big( |12 \rangle - |21 \rangle \big)   ~.
\end{eqnarray}
This can be achieved by moving two atoms, initially in a product state, into the anti-node of the mode of an empty optical cavity. The first atom should enter the cavity in its ground state, while the second one should be prepared in the excited state $|2 \rangle$. Subsequently the atoms are placed in a position where both see the same coupling to the resonator field. 

\subsection{Adiabatic passage}

The entangling process is first described using a simplified model which does not take into account dissipation of the atom-cavity system. In the following, $b$ and $b^\dagger$ denote the creation and annihilation operator for a single photon inside the cavity and $g_i$ is the (real) atom-cavity coupling constant of atom $i$ ($i=1,2$). Then the Hamiltonian describing the interaction between the atoms and the cavity field is given by
\begin{equation} \label{H}
H_{\rm int} = {\rm i}\hbar \sum_{i=1,2} g_i  \, b  \, |2 \rangle_{ii} \langle 1| + {\rm h.c.} 
\end{equation}
Let us denote a state with $n$ photons and the atoms in $|ij\rangle$ in the following as $|ij;n \rangle$. Since the system is initially prepared in a state with only one excitation and the excitation in the system does not change when it develops in time with the Hamiltonian (\ref{H}), the relevant Hilbert space is only three-dimensional and contains the states $|12;0\rangle$, $|21;0\rangle$ and $|11;1\rangle$.

The evolution of the system remains in this subspace and goes by  
\begin{equation} \label{H2}
H_{\rm int} = {\rm i}\hbar  \, \big( g_1 |21;0\rangle + g_2 |12;0\rangle \big) \langle 11;1|  
+ {\rm h.c.}  
\end{equation}
The system possesses one eigenstate with the zero eigenvalue $\lambda_1=0$ given by  
\begin{equation} \label{aesthetic}
|\lambda_1 \rangle= {1 \over R} \, \big( g_1 |12;0 \rangle - g_2 |21;0 \rangle \big) 
\end{equation}
with
\begin{equation} 	\label{v}
R \equiv \big(g_1^2 + g_2^2\big)^{1/2} ~.
\end{equation}
It describes a state with no photon in the cavity mode. Prepared in this state, the atoms cannot transfer their excitation into the resonator due to the effectively vanishing atom-field interaction. Therefore the state (\ref{aesthetic}) can also be called a {\em dark state}. The other two eigenstates are 
\begin{equation}  \label{11}
|\lambda_{2,3} \rangle= {1 \over \sqrt{2} R} \,
\big( g_{2} |12;0 \rangle + g_{1} |21;0 \rangle  \pm  {\rm i} R |11;1 \rangle \big)  
\end{equation}
and correspond to the eigenvalues $\lambda_{2,3} = \mp \hbar R$. 

While the atoms move through the resonator, the atom-cavity coupling constants $g_1$ and $g_2$, and therefore also the eigenstates of the system, change in time. If this happens slowly compared to the time scale given by the eigenvalues $\lambda_{2,3}$, the adiabatic theorem \cite{adia} can be used to predict the time evolution of the system. Atoms that were initially in a dark state remain in a dark state, thereby adiabatically following the change of parameters. Note that this is exactly the case for the state preparation scheme described at the beginning of this section. When the second atom enters the cavity field, at $t=0$ one has $g_1=0$ and $|\lambda_1 \rangle = |12;0 \rangle$. When the atoms reach the anti-node of the quantised standing wave field mode inside the resonator the coupling constants $g_1$ and $g_2$ become the same and the dark state of the system equals $|\lambda_1 \rangle= |a;0 \rangle$. In this situation, both atoms are in a maximally entangled state \cite{remark}.

Let us now analyse this process in more detail with the help of an adiabatic elimination. Since we know already that the system remains in its dark state $|\lambda_1\rangle$  to a very good approximation, it is convenient to decompose its state vector $|\psi \rangle$ as 
\begin{equation} \label{frown}
|\psi(t) \rangle = \sum_{j=1}^3 c_j(t) \, |\lambda_j(t) \rangle ~.
\end{equation}
Using the Schr\"odinger equation and the Hamiltonian (\ref{H}) reveals that the time evolution of the coefficients $c_k$ is governed by the differential equations
\begin{equation} \label{frog}
\dot{c}_k = - {{\rm i} \over \hbar} \, c_k \lambda_k - \sum_{j=1}^3 c_j \, \langle \lambda_k 
| \dot{\lambda}_j \rangle ~.
\end{equation}
Using (\ref{aesthetic})-(\ref{11}), this leads to
\begin{eqnarray} \label{naughty}
\left(  \begin{array}{c} \dot{c_1} \\ \dot{c_2} \\ \dot{c_3} \end{array} \right)
&=& \left(  \begin{array}{ccc} 0 & S & S \\ -S &  {\rm i} R & 0 \\ 
-S & 0 &  -{\rm i} R \end{array} \right) \left(  \begin{array}{c} c_1 \\ c_2 \\ c_3 \end{array} \right)
\end{eqnarray}
with 
\begin{equation}
S \equiv {\dot{g_1}g_2-\dot{g_2} g_1 \over \sqrt{2} \, (g_1^2+g_2^2)} ~. 
\end{equation}
For frequencies $R$ much larger than the frequency $S$ there are two different time scales in the system and the time evolution can be solved by eliminating the fast changing coefficients $c_2$ and $c_3$. To do so it is assumed that they always adapt immediately to the slowly varying coefficient $c_1$ and their derivatives are set equal to zero which yields
\begin{equation} \label{witty}
c_2 = -c_3 = - {{\rm i} S \over R} \, c_1 ~. 		
\end{equation}
Moreover, it can be seen from (\ref{naughty}) that the derivative of $c_1$ equals zero within this approximation.

Provided that the system is initially perfectly prepared in the product state $|12;0\rangle$ one has $c_1(0)=1$ and the state of the system equals in first order in S/R  
\begin{equation} \label{obliged}
|\psi \rangle = {1 \over \| \cdot \|} \,  \big( g_1 |12;0\rangle -g_2  |21;0\rangle  + \sqrt{2}S |11;1 \rangle \big) 
\end{equation}
during the state preparation process. From this the total population in the dark state can be estimated for all times, $F= 1- 2 S^2/R^2$, and differs from one only in second order in $S/R$. Once the atoms stop moving, the derivatives of the coupling constants $g_1$ and $g_2$ and thus also the rate $S$ becomes zero
which yields 
\begin{equation} \label{obliged2}
F(T) = 1~, 
\end{equation}
as expected for an adiabatic process. The fidelity of the finally obtained state does not differ from one and the proposed scheme is very precise. 

As can be seen from (\ref{obliged}), during the whole state preparation, nearly no population accumulates in the cavity mode. Therefore, the scheme also works with a high success rate if cavity decay is taken into account. Let us denote the rate with which a single photon inside the resonator leaks out through the cavity mirrors with $\kappa$. Then the probability for no photon emission equals to a very good approximation 
\begin{equation} \label{p0}
P_0 (T) = 1 - \kappa \int_0^T  {\rm d}t \, \frac{(\dot g_1 g_2 - \dot g_2 g_1)^2} {(g_1^2+g_2^2\, )^3} ~,
\end{equation}
where $T$ is the time it takes to prepare the atoms in the maximally entangled state. The result (\ref{p0}) shows that the photon emission rate in the scheme is proportional to the cavity leakage rate $\kappa$. In case of an emission, the state preparation failed and the experiment has to be repeated. One way to reduce the failure rate by a factor $N$ is to move the atoms $N$ times slower through the resonator. This increases the operation time $T$ by a factor $N$ but decreases the population in the cavity mode by a factor $N^2$. 

\subsection{Dissipation-assisted adiabatic passage}

However, long operation times make the proposed scheme more sensitive to other error sources, like spontaneous emission from the atoms. In the following, leakage of photons is taken into account. First we show that the scheme works as predicted in the previous subsection, even if the decay rate $\kappa$ is about the same size as the maximum atom-cavity coupling constant $g$, as long as the system is operated in the adiabatic regime. But the system can also be operated outside the adiabatic regime. Under the condition of no photon emission, the presence of the cavity decay rate $\kappa$ damps away errors due to non-adiabaticity. Let us assume that photons leaking through the cavity mirrors can be detected with an efficiency close to one. If an emission takes place, the experiment failed and has to be repeated. Otherwise and for a wide range of experimental parameters, the presence of the cavity decay rate has an effect similar to error detection measurements.  

To describe the time evolution of the system in the presence of dissipation we use in the following the quantum jump approach \cite{HeWi11,HeWi2,HeWi3}. It provides a conditional Hamiltonian $H_{\rm cond}$ which describes the time evolution of the system under the condition of {\em no} photon emission. For the system under consideration it equals in the interaction picture with respect to the interaction-free Hamiltonian 
\begin{equation} \label{leap}
H_{\rm cond} = H_{\rm int} - {\textstyle {{\rm i} \over 2}} \hbar \kappa \, b^\dagger b ~.
\end{equation}
This operator can be written as
\begin{eqnarray} \label{H3}
H_{\rm cond} &=& {\rm i}\hbar  \, \big(  g_1  |21;0\rangle + g_2  |12;0\rangle \big) \langle 11;1|  
+ {\rm h.c.} \nonumber \\
&&  - {\textstyle {{\rm i} \over 2}} \hbar \kappa \, |11;1 \rangle \langle 11;1| 
\end{eqnarray}
in the relevant Hilbert space with one excitation in the system. Note that the conditional Hamiltonian is non-Hermitian and the norm of a state vector developing with the corresponding Schr\"odinger equation decreases in general in time. Given the initial state $|\psi \rangle$, the state of the system equals
\begin{eqnarray} 
|\psi^0(T) \rangle &=& U_{\rm cond}(T,0) |\psi \rangle /\| \cdot \|
\end{eqnarray}
at time $T$. For convinience, $H_{\rm cond}$ has been defined such that  
\begin{eqnarray} \label{P0}
P_0(T) &=& \| \, U_{\rm cond}(T,0) |\psi \rangle \, \|^2
\end{eqnarray}
equals the probability for no photon in $T$.

Calculating the eigenvalues and eigenstates of the conditional Hamiltonian (\ref{H3}) one finds that the system still possesses a zero eigenvalue $\lambda_1=0$ in the presence of a finite cavity decay rate $\kappa$. The dark state $|\lambda_1 \rangle$ of the system is the same as for $\kappa=0$ and is given in (\ref{aesthetic}). Only the eigenvalues $\lambda_{2,3}$ and the eigenstates $|\lambda_{2,3} \rangle$ change and and are given by  
\begin{equation} \label{hogwash}
\lambda_{2,3} = - {\textstyle {{\rm i} \over 4}} \hbar \kappa \mp \hbar \,  \big( g_1^2 + g_2^2 
- {\textstyle {1 \over 16}} \kappa^2 \big)^{1/2} ~.
\end{equation}
As long as the atoms move such that the parameters $g_1$ and $g_2$ change slowly on the time scale given by the real parts of  the eigenvalues $\lambda_{2,3}$, the no-photon time evolution of the atoms inside the cavity can again be predicted by the adiabatic theorem. As expected, the system remains in the dark state $|\lambda_1 \rangle$ to a very good approximation.

To calculate the probability and fidelity of the prepared state we proceed as in the previous subsection and adiabatically eliminate the fast varying amplitudes of the state vector. Due to the non-Hermiticity of the conditional Hamiltonian,  the eigenstates $|\lambda_{2,3} \rangle$ are no longer orthogonal to each other; for certain parameters of $g_i$ and $\kappa$ they can even become degenerate. It is therefore more convinient to consider the basis vectors 
\begin{eqnarray} \label{aisle}
&& \hspace{-1cm} 
|\eta_1 \rangle \equiv {1 \over R} \, \big( g_1 |12;0 \rangle - g_2 |21;0 \rangle \big) = |\lambda_1 \rangle ~,
\nonumber \\
&&  \hspace{-1cm}  |\eta_2 \rangle \equiv {1 \over R} \, \big( g_2 |12;0 \rangle + g_1 |21;0 \rangle \big) ~,~
|\eta_3 \rangle \equiv |11;1 \rangle 
\end{eqnarray}
and to define
\begin{eqnarray} \label{aisle2}
|\psi(t) \rangle = \sum_{j=1}^3 c_j(t) \, |\eta_j(t) \rangle ~.
\end{eqnarray}
Proceeding as in the previous subsection leads to
\begin{equation} \label{frog2}
\dot{c}_k = - \sum_{j=1}^3 c_j \, \left(  {{\rm i} \over \hbar} \, \langle \eta_k| H_{\rm cond} | \eta_j \rangle
+ \langle \eta_k | \dot{\eta}_j \rangle \right) 
\end{equation}
which gives the differential equations
\begin{eqnarray} \label{naughty2}
\left(  \begin{array}{c} \dot{c_1} \\ \dot{c_2} \\ \dot{c_3} \end{array} \right)
&=& \left(  \begin{array}{ccc} 0 & \sqrt{2} S & 0 \\ -\sqrt{2} S & 0 & R \\ 
0 & -R & - {\textstyle {1 \over 2}} \kappa \end{array} \right) 
\left(  \begin{array}{c} c_1 \\ c_2 \\ c_3 \end{array} \right) ~. 
\end{eqnarray}
If $S$ is much smaller than $R$, this differential equation can be solved by setting the derivatives of the fast varying coefficients $c_2$ and $c_3$ equal to zero. This leads to 
\begin{eqnarray} \label{naughty3}
c_2 = -{\kappa S \over \sqrt{2} R^2} \, c_1 ~,~ c_3 = -{\sqrt{2} S \over R} \, c_1  ~.
\end{eqnarray}
Again it can be shown that only a small population proportional $S^2$ accumulates outside the dark state $|\eta_1 \rangle = |\lambda_1 \rangle$ during the adiabatic population transfer. When the atoms stop at the end of the state preparation, the rate $S$ becomes zero and it is $c_{2,3}(T)=0$. From this one sees immediately that the fidelity of the finally obtained (normalised) state is again one,  
\begin{equation} \label{xxx1}
F(T) = 1 ~,
\end{equation}
in case of adiabaticity. Under the condition of no emission, the system moves into $|\lambda_1(T) \rangle$. 

Different from the case with $\kappa=0$, the derivative of the amplitude $c_1$ is now no longer negligible and it can be shown that 
\begin{eqnarray} \label{pain}
\dot{c}_1 &=& - {\kappa S^2 \over R^2} \,  c_1  ~.
\end{eqnarray}
Solving this differential equation for the initial condition $c_1(0)=1$ leads to
\begin{equation} \label{xxx}
c_1(T) =  \exp \left( - \kappa \int_0^T {\rm d}t \, {S^2  \over R^2} \right) ~. 
\end{equation}
This is the amplitude of the dark state with respect to the unnormalised state of the system at time $t$ and under the condition of no photon emission. The success rate of the state preparation can be obtained from (\ref{P0}) and equals $|c_1(T)|^2$ since all population is in the state $|\lambda_1 \rangle$. Using (\ref{xxx}) this leads to
\begin{equation} \label{xxx2}
P_0(T) = \exp \left(  - \kappa \int_0^T {\rm d}t \, 
\frac{(\dot g_1 g_2 - \dot g_2 g_1)^2} {(g_1^2+g_2^2\, )^3} \right) 
\end{equation}
which agrees with the result given in (\ref{p0}) up to first order in $\kappa/R$. As long as the atoms move slowly enough into the cavity and the scheme is operated in the {\em adiabatic regime}, the system behaves as predicted in the previous subsection. 

However, the presence of a finite decay rate $\kappa$ can have a dramatic effect on the time evolution if the system is operated {\em outside} the adiabatic regime. We show in the following that the state preparation time $T$ can be shortened by as much as two orders of magnitude by moving the atoms with higher velocities into the resonator. Under the condition of no photon emission the system behaves as predicted by the adiabatic theorem 
even if the adiabaticity condition $S \ll R$ is no longer fulfilled. The price one has to pay for the speed up of the scheme is that now photon emission might occur during the state preparation. However, the success rate of the proposed scheme is about $90 \, \%$ for a wide range of parameters. If photons are detected with a high efficiency, the experiment can be repeated if necessary. 

The main effect of the finite cavity decay rate $\kappa$ on the no-photon time evolution of the system is that the eigenvalues $\lambda_{2,3}$ accumulate an imaginary part proportional to $\kappa$ (see (\ref{hogwash})). These imaginary parts result from the non-Hermitian terms in the conditional Hamiltonian and damp away the excitation in unwanted states, namely in the eigenstates $|\lambda_{2,3} \rangle$, as a consequence of no-photon measurements. This leads to a decrease of the norm of the state vector of the system and to a finite success rate of the state preparation but also increases the fidelity of the finally prepared state. To a very good approximation, dissipation has the same effect as error detection measurements of whether the system behaves as predicted for the adiabatic regime or not. The time evolution of the system becomes a dissipation-assisted adiabatic passage.

The basic ingredient for dissipation-assisted adiabatic passages is the existence of de\-co\-he\-ren\-ce-free (DF) states. In the example considered here the only DF state of the system with respect to cavity decay and with one quantum of excitation is the dark state $|\lambda_1 \rangle$. Populating this state cannot lead to a photon emission. Which states of a system are DF depends on the system parameters. If these parameters change in time, a time evolution can be induced during which the system remains DF. It adiabatically follows the changing parameters. The idea underlying the proposed state preparation scheme can easily be carried over to other setups \cite{simple}.

To investigate the influence of the cavity decay rate $\kappa$ in the non-adiabatic regime numerically, we consider in the following an optical resonator with a Gaussian mode profile and denote the cavity waist by $w_0$. As an example, it is assumed that atom 1 is initially placed about four waist lengths away from the cavity centre, $x_1(0) =-4 \, w_0$, while the second is placed at $x_2(0)=-6 \, w_0$. As a function of the atom position, the atom-cavity coupling constant equals 
\begin{equation} \label{pigeon}
g_i (x_i) = g \, \exp \big( - (x_i/w_0)^2 \big)  ~,
\end{equation}
where $g$ is the maximum coupling rate assumed only at an anti node of the field mode inside the resonator. In the following we consider the case where both atoms move with the same velocity and as shown in Figure \ref{setup}(a). To asssure that the atoms stop at the position where both see the same coupling constant, i.e.~$x_1(T)=w_0$ and $x_2(T)= - w_0$, we choose
\begin{equation} \label{velo}
v_1 = v_2 = v_{\rm max} \, \sin^2 \big( \pi (x_1+4 \, w_0)/5 \, w_0 \big) ~.
\end{equation}
Because $v_1(T)=v_2(T)=0$, deviations of the fidelity of the finally prepared state from one are only expected in case of relatively high velocities $v_{\rm max}$ due to the non-adiabaticity of the time evolution. 

\begin{figure}
\begin{minipage}{\columnwidth}
\begin{center}
\resizebox{\columnwidth}{!}{\rotatebox{0}{\includegraphics{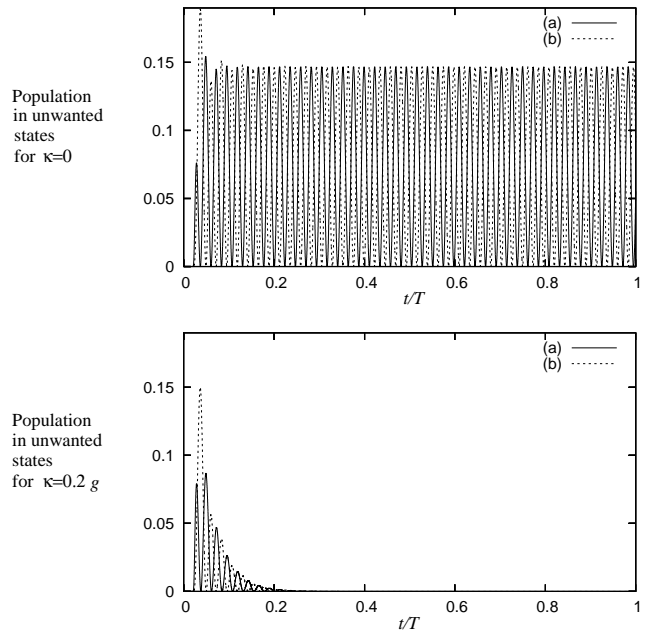}}}  
\end{center}
\vspace*{-0.7cm}
\caption{Population in the states $\big( g_2\, |12;0 \rangle +  g_1 \, |21;0 \rangle \big)/R$ (a) and $|11;1 \rangle$ (b) 
as a function of the time $t$ and for $\kappa=0$ and $\kappa = 0.2 \, g$ and for  $v_{\rm max} = 5 \, w_0g$.} \label{move3}
\end{minipage}
\end{figure}

Figure \ref{move3} results from a numerical solution of the time evolution of the system with the Hamiltonian (\ref{leap}). If the velocity $v_{\rm max}$ in (\ref{velo}) becomes too large and $\kappa=0$, the fidelity of the prepared state equals no longer one. Population that accumulates during the state preparation process in non-dark states performs Rabi oscillations between the atoms and the cavity mode and returns no longer into the dark state $|\lambda_1 \rangle$ when the atoms come to rest and $S$ becomes zero (see upper graph in Figure \ref{move3}). The presence of a cavity leakage rate $\kappa$ helps achieving fidelities close to one in the non-adiabatic regime by damping away the population in non-dark states (see lower graph in Figure \ref{move3}). 

\begin{figure}
\begin{minipage}{\columnwidth}
\begin{center}
\resizebox{\columnwidth}{!}{\rotatebox{0}{\includegraphics{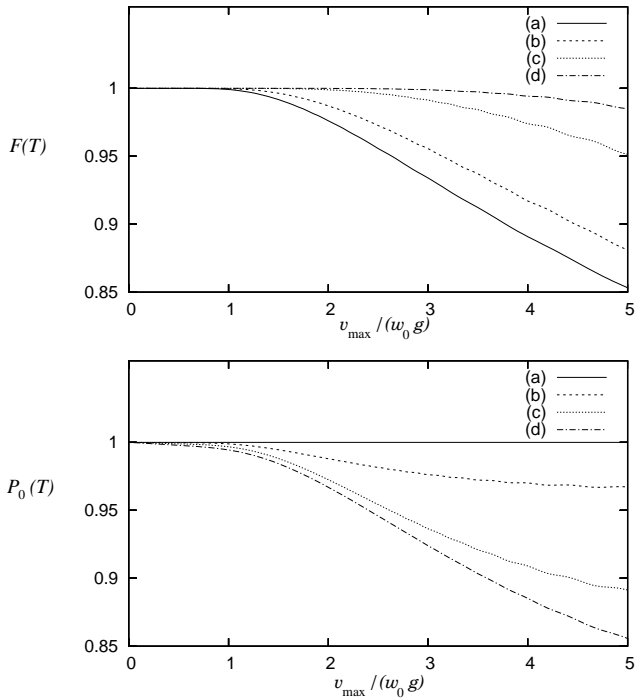}}}  
\end{center}
\vspace*{-0.7cm}
\caption{Fidelity and success rate of the prepared state as a function of the atom velocity $v_{\rm max}$ and for the cavity decay rates $\kappa=0$  (a), $\kappa=0.01 \, g$ (b), $\kappa=0.05 \, g$ (c) and $\kappa=0.1 \, g$ (d).} \label{move2}
\end{minipage}
\end{figure}

The fidelity of the finally prepared state as a function of the atom velocity $v_{\rm max}$ is shown in Figure \ref{move2}. As expected, the fidelity decreases the faster the atoms move into the cavity. The presence of a higher cavity decay rate improves the fidelity while the gate failure rate due to photon emission increases by about the same amount. For decay rates $\kappa$ about one order of magnitude smaller than the atom-cavity coupling constant $g$, the possibility of a photon emission has the same effect as an error detection measurement. 

\begin{figure}
\begin{minipage}{\columnwidth}
\begin{center}
\resizebox{\columnwidth}{!}{\rotatebox{0}{\includegraphics{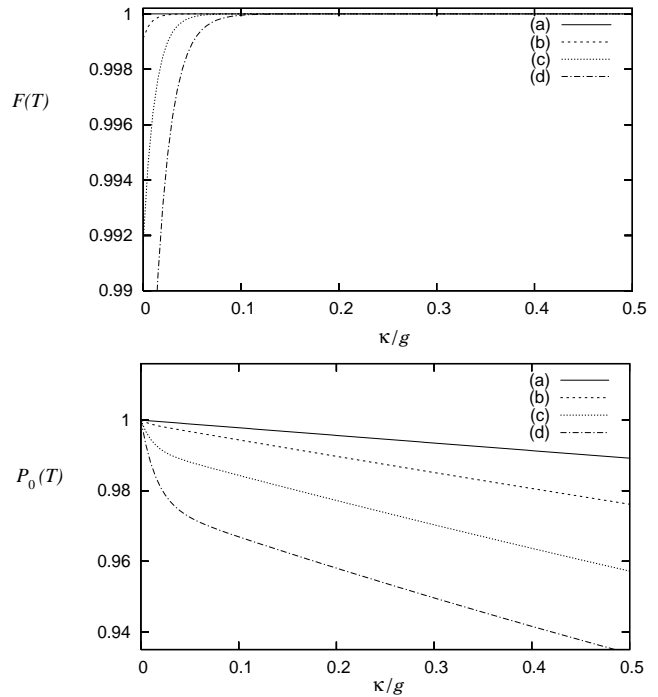}}}  
\end{center}
\vspace*{-0.7cm}
\caption{Fidelity and success rate as a function of the cavity decay rate $\kappa$ for $v_{\rm max}=0.5 \, w_0 g$ (a), $v_{\rm max}=w_0 g$ (b), $v_{\rm max}=1.5 \, w_0 g$ (c) and $v_{\rm max}=2 \, w_0 g$ (d).} \label{move4}
\end{minipage}
\end{figure}

Depending on the atom velocity $v_{\rm max}$ one can allow relatively large cavity leakage rates $\kappa$ in the scheme. If $\kappa$ is larger than in the case where its presence has the same effect as a perfect error detection measurement,  the success rate of the state preparation decreases below the fidelity of the scheme in case of $\kappa=0$ (see Figure \ref{move4}). In the parameter regime (\ref{aim}), considered in this section, the fidelity of the finally obtained state does not differ from one.

\subsection{State preparation using the inverse quantum Zeno effect}

For a wide parameter regime, the proposed state preparation scheme can alternatively be understood as a concrete realisation of the inverse quantum Zeno effect. Similar to the original quantum Zeno effect \cite{misra,behe}, the inverse Zeno effect \cite{aharonov} predicts the time evolution of a system on which rapidly repeated measurements are performed, though these measurements vary in time and it is measured  whether the time evolution of a system follows a certain assumed trajectory. If the time between subsequent measurements tends to zero, the state vector of the system follows the assumed trajectory even if there is no interaction inducing a time evolution in the system. In this subsection we shortly comment on this point of view.

Given fixed values for $g_1$ and $g_2$, the interaction of the atom-cavity system with its environment over a time $\Delta t$ can be shown to have the same effect as an ideal measurement whether the system is prepared in the dark state $|\lambda_1 \rangle$ or not, if $\Delta t$ exceeds a certain minimum length. One possibility to show this is to proceed as in \cite{beige4} and to calculate the no-photon time evolution of the system over $\Delta t$ using the Hamiltonian (\ref{H3}). In this way it can be shown that $U_{\rm cond}(\Delta t,0)$ equals the projector onto the dark state of the system, $|\lambda_1 \rangle \langle \lambda_1|$. The dark state $|\lambda_1 \rangle$ is a decoherence-free state with respect to spontaneous emission through the cavity mirrors and seeing no photon over a time $\Delta t$ assures that the system is in a decoherence-free state. 

To be able to use the inverse quantum Zeno effect to predict the time evolution of the system, the atom cavity coupling constants $g_1$ and $g_2$ should change very slow so that they can be considered as constant over a time $\Delta t$. From one time interval to the other, the measurement performed on the system changes and the system gets projected onto the dark state $|\lambda_1 \rangle$ defined by the actual values of $g_1$ and $g_2$. The inverse quantum Zeno effect implies that the probability to find the system always in the dark (or decoherence-free) state is the closer to one the slower the atoms move inside the cavity. This is in good agreement with the analytical results (\ref{xxx1}) and (\ref{xxx2}) obtained from an adiabatic elimination of the fast varying amplitudes of the state vector. 

The alternative interpretation of the time evolution of the system in the presence of dissipation given in this subsection might help to understand intuitively why success rate and fidelity of the proposed state preparation scheme are so close to one even if $\kappa$ is about the same size as the maximum atom-cavity coupling constant $g$. Schemes based on quantum Zeno effects or adiabatic passages are in general very efficient to suppress one source of dissipation in a system and to create very simple schemes for the preparation of entangled states or the realisation of quantum gates. Nevertheless, they are very slow. To increase the robustness of a scheme with respect to different sources of dissipation, like cavity leakage {\em and} spontaneous emission from the atoms, one should use dissipation-assisted adiabatic passages and operate the system outside the adiabatic regime.

\section{Experimental realisation of maximally entangled two-atom states} \label{section}

\begin{figure}[h]
\begin{minipage}{\columnwidth}
\begin{center}
\resizebox{\columnwidth}{!}{\includegraphics{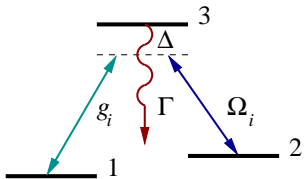}} 
\end{center}
\vspace*{-0.5cm}
\caption{Level configuration of atom $i$. The Rabi frequency of the laser field addressing the $2-3$ transition of atom $i$ is denoted by $\Omega_i$ and $g_i$ is the atom-cavity coupling constant with respect to the $1-3$ transition. Both transitions should have the same detuning $\Delta$ while $\Gamma$ is the spontaneous decay rate of level 3.} \label{fig2}
\end{minipage}
\end{figure}

In this section we discuss the limitations of entangled state preparation via dissipation-assisted adiabatic passages with respect to dissipation, including spontaneous emission from the atoms. While cavity decay can even be used to remedy errors, spontaneous decay of excited atomic levels remains the main obstacle in the proposed state preparation scheme. To assure that the finally obtained state is stable with respect to spontaneous emission from the atoms, the states $|1 \rangle$ and $|2 \rangle$ should be obtained from two different ground states of the atom. To couple these two levels an additional atomic level $3$ and a coupling laser focussed on the region of the cavity can be used, as shown in Figure \ref{fig2}.  

In the following, $\Omega_i$ denotes the Rabi frequency and $\Delta$ the detuning with which the coupling laser excites the $2-3$ transition in atom $i$. Note that the scheme does not require individual addressing of the atoms inside the cavity. The index $i$ accounts for the possible dependency of the Rabi frequency on the atomic position. 
In the following we solve the no-photon time evolution of the system numerically and assume as a concrete example that 
\begin{equation} \label{pigeon2}
\Omega_i (x_i) = g \, \exp \big( - (x_i/(5w_0))^2 \big)  ~.
\end{equation}
The waist of the laser field is five times the cavity waist $w_0$ defined in (\ref{pigeon}). 

Again we denote the atom-cavity coupling constant of atom $i$, now with respect to the $1-3$ transition, by $g_i$
and assume a Gaussian mode profile as in (\ref{pigeon}). The detuning of the cavity mode should be the same as the detuning $\Delta$ of the coupling laser. Assuming $N$ atoms in the cavity, the conditional Hamiltonian equals
\begin{eqnarray} \label{conceal}
H_{\rm cond} &=& \hbar \sum_{i=1}^N  {\textstyle {1 \over 2}} \Omega_i \, |2 \rangle_{ii} \langle 3| 
+ {\rm i} g_i b \, |3 \rangle_{ii} \langle 1| + {\rm h.c.} \nonumber \\
&& + \hbar (\Delta - {\textstyle {{\rm i} \over 2}} \Gamma)  \sum_{i=1}^N |3 \rangle_{ii} \langle 3|  
- {\textstyle {{\rm i} \over 2}} \hbar \kappa \, b^{\dagger}b
\end{eqnarray}
in the interaction picture with respect to the interaction-free Hamiltonian minus $\sum_{i=1}^N \hbar \Delta |3 \rangle_{ii} \langle 3|$ and taking both types of dissipation into account. 

To assure that the atoms behave like two-level atoms the detuning $\Delta$ should be at least ten times larger than the system parameters $g_i$, $\Omega_i$ and $\Gamma$. This allows for an adiabatic elimination of  level 3
resulting in the effective conditional Hamiltonian
\begin{eqnarray} \label{literal}
H_{\rm cond} &=& {\rm i} \hbar \sum_{i=1}^N  \tilde{g}_i b \, |2 \rangle_{ii} \langle 1|  + {\rm h.c.} 
- {\textstyle {{\rm i} \over 2}} \hbar \, \tilde{\Gamma}_i \sum_{i=1}^N |2 \rangle_{ii} \langle 2|  \nonumber \\
&& - {\textstyle {{\rm i} \over 2}} \hbar \tilde{\kappa} \, b^{\dagger}b ~.
\end{eqnarray}
The atom-cavity coupling constant of the reduced level scheme equals 
\begin{equation} \label{obstruct}
\tilde{g}_i = {\Omega_i \over 2\Delta} \, g_i
\end{equation}
in first order in $1/\Delta$. Within this approximation, spontaneous decay from the atoms is negligible.  However, since the operation time of the scheme is larger than the inverse atom-cavity coupling constants, higher order corrections have to be taken into account. Doing so leads to the effective spontaneous decay rate 
\begin{equation} \label{geek}
\tilde{\Gamma}_i = \left( {\Omega_i \over 2\Delta} \right)^2  \Gamma
\end{equation}
assigned to level $2$ in (\ref{literal}). While changing the atom-cavity coupling and the atom decay rate, the detuning has no effect on the cavity leakage rate and it is
\begin{equation} \label{geek2}
\tilde{\kappa} = \kappa ~.
\end{equation}
Note that when the rates $g$, $\Gamma$ and $\kappa$ are about the same size, as in (\ref{aim}), the effective rates follow the ordering $\tilde{\kappa} \gg \tilde g_i \gg \tilde \Gamma_i$. 

Since the presence of the detuning increases the relative cavity decay rate $\tilde \kappa / \tilde g$ (with $\tilde g \equiv {\rm max} \, \tilde g_i$) significantly, the system is now no longer operated in a parameter regime where the atom-cavity coupling constant is effectively of similar size as the decay rates, even if $g \sim \kappa \sim \Gamma$. Therefore, requesting a certain minimum fidelity and success rate in the presence of Raman transitions, one can only allow a relatively small amount of dissipation in the system (for comparison see \cite{beige3,jane}). Concrete numerical results are presented in  the next two subsections. 

\subsection{A two-atom scheme} \label{two}

Apart from the finally prepared state being stable, another advantage of using $\Lambda$-systems is that this allows for a simplification of the state preparation scheme discussed in Section \ref{sectio}. By turning off the laser field, the coupling of the atoms to the field mode can be interupted whenever the atoms reach a position where they should no longer interact with the cavity field. Instead of having to move the atoms exactly into a certain position they can move with constant velocity through the resonator. When both see the same atom-cavity coupling constant, the laser field is turned off and the state of the atoms changes no longer in time. Afterwards the atoms can be moved out of the resonator without destroying the maximally entangled state. 

\begin{figure}
\begin{minipage}{\columnwidth}
\begin{center}
\resizebox{\columnwidth}{!}{\rotatebox{0}{\includegraphics{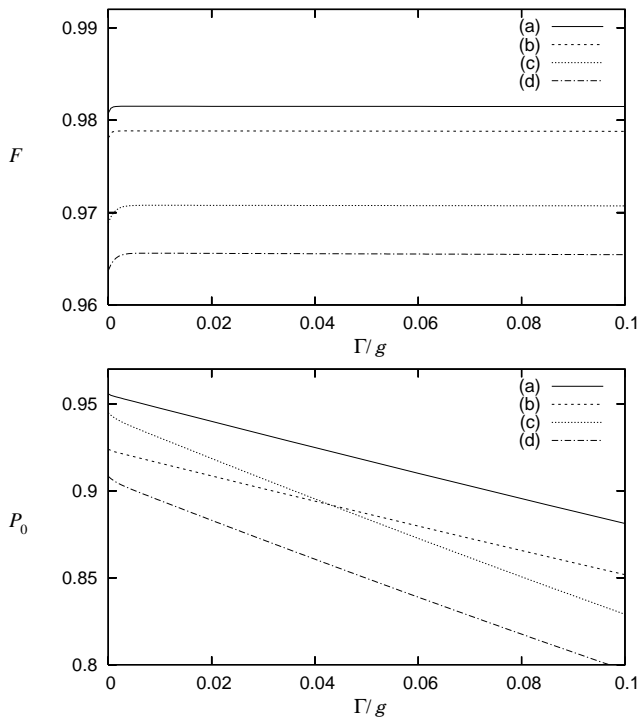}}}  
\end{center}
\vspace*{-0.7cm}
\caption{Fidelity and success rate of a state preparation scheme where two atoms move with constant speed through the resonator as a function of the atom decay rate $\Gamma$ for 
$v=0.002 \, w_0g$,  $\Delta=20 \, g$ and $\kappa= \tilde g = 0.025 \, g$ (a), 
$v=0.002 \, w_0g$,  $\Delta=20 \, g$ and $\kappa= 2 \, \tilde g = 0.05 \, g$ (b), 
$v=0.005 \, w_0g$,  $\Delta=10 \, g$ and $\kappa= \tilde g = 0.05 \, g$ (c) and 
$v=0.005 \, w_0g$,  $\Delta=10 \, g$ and $\kappa= 2 \tilde g = 0.1 \, g$ (d). 
The distance of the atoms equals one cavity waist $w_0$.} \label{move7}
\end{minipage}
\end{figure}

Again the system should initially be prepared in the state $|12;0\rangle$ with the atoms outside the cavity. 
Figure \ref{move7} shows the fidelity under the condition of no photon emission and the success rate of the state preparation after the atoms left the resonator as a function of the atom decay rate $\Gamma$. Increasing $\Gamma$ leads to a decrease of the no-photon emission probability. As in the previous section, it is assumed that photons can be detected with a high efficiency and the experiment is repeated whenever necessary. If the detuning $\Delta$ is much larger than $g$ then most photons result from leakage through the cavity mirrors and not from the atoms since $\tilde \kappa \gg \tilde \Gamma_i$. 

Numerical simulations show that the presence of a decay rate $\kappa$ of the order of the effective atom-cavity coupling constant $\tilde g$ can indeed increase the fildelity of the finally prepared state compared to the case where $\kappa=0$. For larger values of $\kappa$, like $\kappa = 2 \, \tilde g$, the fidelity decreases again. To obtain fidelities close to one in this case, the atoms have to move slowly through the resonator and the system has to be operated closer to quantum Zeno effect regime. The corresponding long state preparation time then leads to a decerease of the spontaneous decay rate $\Gamma$ that can be allowed in the system. No photon probabilities and success rates around $80 \, \%$ can be achieved for $g^2 \sim 100 \, \kappa \Gamma$ (see Figure  \ref{move7}(d)) while $P_0 > 85 \, \%$ requires $g^2 \sim 200 \, \kappa \Gamma$  (see Figure  \ref{move7}(b)). With respect to dissipation, the proposed state preparation scheme is comparable with other atom-cavity schemes \cite{sorensen,pachos,you,pellizzari} while the process itself is much simpler.

Note that it is always possible to obtain fidelities equal to one. This is achieved if the atoms rest for a short time in the position where they both see the same atom-cavity coupling before the laser field with Rabi frequency $\Omega$ is turned off. Then the population still left in unwanted states at the end of the operation can be damped away so resulting in the preparation of the antisymmetric state $|a \rangle$ as described in \cite{plenio}. 

\subsection{A three-atom scheme} \label{tree}

A further improvement of the feasibility of the proposed experiment can be obtained from a straightforward generalisation of the state preparation scheme to the three-atom case. The main advantage of using three atoms is that the scheme no longer requires to turn off the laser field when the atoms reach a certain position in the cavity. Systems with three atoms in the cavity possess a three-dimensional DF subspace spanned by the ground state $|111;0 \rangle$ and two states with one excitation in the atomic state $|2 \rangle$. Proceedings as in \cite{beige4}, the two other states can be found by orthogonalising  the states
\begin{eqnarray}
&& |\eta_{12} \rangle \equiv {1 \over \| \cdot \|} \, \big( g_1 |121;0 \rangle - g_2 |211;0 \rangle \big) ~, \nonumber \\
&& |\eta_{13} \rangle \equiv {1 \over \| \cdot \|} \, \big( g_1 |112;0 \rangle - g_3 |211;0 \rangle \big) ~, \nonumber \\
&& |\eta_{23} \rangle \equiv {1 \over \| \cdot \|} \, \big( g_2 |112;0 \rangle - g_3 |121;0 \rangle \big) ~,
\end{eqnarray}
which can easily be identified as DF states. For more than two atoms there are in general several states with the same amount of excitation in the atoms and it is more difficult to predict the outcome of the state preparation scheme than in the two-atom case, where one can easily deduce the final state from the fact that the amount of excitation in the system does not change.

\begin{figure}
\begin{minipage}{\columnwidth}
\begin{center}
\resizebox{\columnwidth}{!}{\rotatebox{0}{\includegraphics{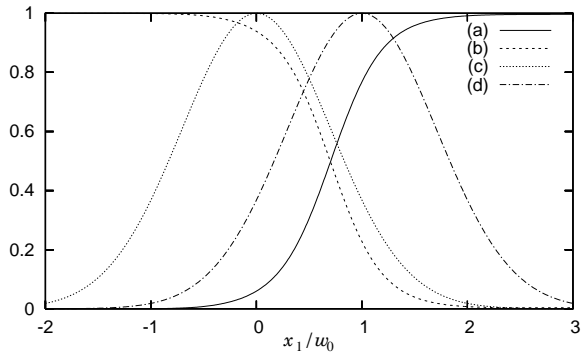}}}  
\end{center}
\vspace*{-0.7cm}
\caption{Population in the entangled state $|s1;0 \rangle$ (a) and in the initial state $|112;0\rangle$ (b) and the atom-cavity coupling constants $g_1=g_2$ (c) and $g_3$ (d) as a function of the position $x_1$ of atom 1 and 2. The distance of the first two atoms from the third equals one cavity waist $w_0$. They move through the cavity with constant speed $v=0.002 \, w_0g$ while $\Delta = 20 \, g$, $\kappa= 0.02 \, g$ and $\Gamma = 0.05 \, g$. If no photon is emitted the maximally entangled state (\ref{s}) of atom 1 and 2 is prepared with a fidelity of $F=99.7 \, \%$ and a success rate of $87.6 \, \%$.} \label{move8}
\end{minipage}
\end{figure}

As an example, let us consider a simple scheme using three atoms and aiming at the preparation of the maximally entangled symmetric state
\begin{eqnarray} \label{s}
|s \rangle & \equiv & {1 \over \sqrt{2}} \, \big( |12 \rangle + |21 \rangle \big)  
\end{eqnarray}
of two atoms. This can be achieved by moving the atoms with constant speed $v$ through the cavity using a setup similar to the one shown in Figure \ref{setup}(b). Atom 1 and 2 should enter the cavity in the ground state $|1 \rangle$ and see all the time the same cavity coupling constant. This can be achieved by moving the atoms parallel through an anti-node of the cavity; alternatively a ring cavity could be used. The third atom should initially be prepared in $|2 \rangle$ and enters the cavity a bit later but in a way that all three atoms interact at some point simultaneously with the cavity mode. 
  
Initially prepared in the state $|112\rangle$, the atoms enter the cavity in a DF state. The state $|112;0\rangle$ of the system is a superposition of the states $ |\eta_{13} \rangle$ and $ |\eta_{23} \rangle$ and can be written as
\begin{eqnarray} \label{sup}
&& \hspace*{-0.5cm} {1 \over \sqrt{2}} \, \big( |\eta_{13} \rangle + |\eta_{23} \rangle \big) = \nonumber \\
&& {1 \over \big( (g_1^2+g_2^2) + 2 g_3^2 \big)^{1/2}} \, 
\big( (g_1+g_2) |112;0 \rangle - \sqrt{2} g_3 |s1;0 \rangle \big) \nonumber \\ &&
\end{eqnarray}
with $g_3=0$. When atom 1 and 2 leave the cavity, the third atom is in the ground state since this is the only DF state with only one particle in the resonator. After all atoms passed through the cavity, the first two atoms are in the maximally entangled symmetric state $|s \rangle$. They now equally share the excitation initially in atom 3. In the setup considered here, the system remains continuously in the superposition (\ref{sup}) and the final state of the atoms is reached when $g_1=g_2=0$ \cite{carsten}. 

Figure \ref{move8} shows the population in the states $|112;0 \rangle$ and $|s1;0 \rangle$ as a function of the 
position of atom 1 and 2 in the cavity mode and results from a numerical integration of the Schr\"odinger equation given by (\ref{conceal}). Choosing the experimental parameters similar to the parameters in Figure \ref{move7}, it is found that atom 1 and 2 leave the cavity indeed in a maximally entangled state.

\section{Conclusions} \label{conc}

In this paper we discussed state preparation schemes aiming at the creation of a maximally entangled state of two two-level atoms. This can be achieved by moving either two or three atoms, initally prepared in a non entangled state, with constant speed through an optical cavity. In the two-atom case, the first atom enters the resonator in its ground state while the second atom is initially prepared in the excited state. When both atoms reach a position where both see the same cavity coupling, the interaction with the resonator mode is turned off. This is possible when the atom-cavity interaction is established indirectly via an auxiliary level and with the help of a laser field. Individual laser addressing of atoms inside the cavity is not required.

To further improve the feasibility of the state preparation it has been proposed to use three atoms. Again the atoms move with constant speed through the resonator. The first two atoms enter the cavity in the ground state such that they always see the same coupling to the resonator mode. If the third atom is initially prepared in the excited state and enters the cavity region shortly after the others, then atom 1 and 2 leave the resonator in the maximally entangled symmetric state. Different from the two-atom case, the three-atom scheme no longer depends on the accuracy with which the coupling laser can be turned off at the right moment. It is sufficient to focus the laser on the region of the cavity and the scheme does not require precise control of the experimental parameters.

The basic mechanism underlying the proposed state preparation schemes is that the atoms enter the cavity in an eigenstate of the atom-cavity interaction Hamiltonian. When the atoms move through the cavity, the atom-cavity coupling and eigenstates of the system change and a time evolution is induced. The system follows the changing parameters adiabatically and remains in an eigenstate. 

Other advantages of the scheme result from the fact that the only populated eigenstates in the scheme are the zero eigenstates of the atom-cavity interaction Hamiltonian and therefore the decoherence-free states of the system with respect to cavity decay. Because of this, the scheme can be implemented in the presence of relatively high decay rates $\kappa$. Intuition suggests that dissipation is always damaging. Contrary to this, the presence of a cavity leakage rate allows here to operate the system faster than in the adiabatic regime. Dissipation acts like an error detection measurement and stabilises the desired time evolution by damping away population in unwanted states. Since the time evolution of the system is as expected for an adiabatic process, it can be called a dissipation-assisted adiabatic passage.  

Like in other STIRAP processes \cite{bergmann,berg2},  the fidelity of the finally prepared state depends only on the experimental parameters at the end of the preparation process and the proposed scheme is relatively robust against parameter fluctuations.  For example, in the two-atom case the fidelity of the atomic state depends only on the size of the atom-cavity coupling constants $g_1$ and $g_2$ at the time when the laser field is turned off and the atom-cavity interaction is interrupted (see Section III). However, the parameters at the end of the operation have to be controlled well. If $g_1$ and $g_2$ are not the same, the atoms are prepared in the state $(g_1 |12 \rangle - g_2 |21 \rangle)/\| \cdot \|$ which overlaps with the maximally entangled state with the fidelity $F= {1 \over 2} + g_1g_2/(g_1^2+g_2^2)$. 

A disadvantage of schemes based on dis\-si\-pa\-tion-assisted adiabatic passages is that, when they are operated outside the adiabatic regime, the success rate of the scheme decreases. Photons might be emitted resulting in a failure of the state preparation. If the loss of photons is mainly caused by leakage of photons through the resonator mirrors, this can be detected  with a high efficiency and the experiment can be repeated if necessary. The fidelity of the finally prepared state under the condition of no photon emission is well above $95 \,\%$ for a wide range of experimental parameters. Because the state preparation time of the scheme can be relatively short, success rates above $80 \, \%$ can be achieved for $g^2=100 \, \kappa \Gamma$ (see Figure \ref{move7}(d)) while $P_0 > 85 \, \%$ requires $g^2 \sim 200 \, \kappa \Gamma$ (see Figure  \ref{move7}(b)). With respect to the dissipation problem, the scheme is comparable to other atom-cavity schemes \cite{sorensen,pachos,you,pellizzari}.

A straightforward generalisation of the state preparation scheme discussed here is the preparation of $N$ atoms in a so-called {\em W} state \cite{duer}.  Main characteristics of {\em W} states is that {\em all} atoms share one excitation. Like Bell states, they are highly entangled but their entanglement is more robust. A state measurement on one of the atoms leads only to a relatively small decrease of the entanglement in the system. Hence {\em W} states are a crucial ingredient for optimal cloning protocols \cite{buzek,murao,bruss,simon}. To prepare a {\em W} state, the atoms should initially be prepared in a state with only one of them excited. The first atom has to enter the cavity in the ground state. Beside that, there are no conditions on the state in which the other atoms enter the cavity. For other schemes aiming at the preparation of {\em W} states in atom-cavity systems see \cite{wang,guo,guo2,foeldi}. 

In addition, the proposed scheme can also be generalised to higher excited states. The atoms can be moved all together into the cavity field, some of them could be trapped at fixed positions between the mirrors, as shown in Figure \ref{setup}(b), but they can also be moved repeatedly in and out of the resonator field. To avoid photon emission it is crucial that the atoms enter the resonator field in a decoherence-free (or dark) state. One application of $N$-atom state preparation schemes is adiabatic quantum computation \cite{farhi}. 

More general, dissipation-assisted adiabatic time evolutions can be used in many setups to induce a time evolution inside a decoherence-free subspace by simply changing the experimental parameters that define its states. This idea leads to time evolutions that are widely independent from the exact values of experimental parameters and relatively robust with respect to dissipation. 

\vspace*{0.5cm} 
\noindent {\em Acknowledgment}. We thank the European Union Network of Excellence, QUIPROCONE (project 43), for support of a research visit of C.M. at Imperial College London and A.B. thanks the Royal Society for funding as a University Research Fellow. This project was supported in part also by the EPSRC.


\begin{references}
\bibitem{remark}
If the atoms are then moved out of the resonator on the other side, the dark state of the system becomes $|\lambda_1 \rangle= |21;0 \rangle$ and the entanglement between the atoms would be lost again. How this can be avoided is described in Section \ref{section}.
\end{references}
\end{document}